\documentclass[12pt, final]{l4dc2022}

\usepackage{graphicx}
\usepackage{nicefrac}
\usepackage{mathtools}
\usepackage{hyperref}

\newtheorem{assumption}{Assumption}

\usepackage{algorithm}
\usepackage{algorithmic}

\title[Learning-based Moving Horizon Estimation]{Learning-based Moving Horizon Estimation through Differentiable Convex Optimization Layers}
\usepackage{times}

\author{%
 \Name{Simon Muntwiler} \Email{simonmu@ethz.ch}\\
 \Name{Kim P. Wabersich} \Email{wkim@ethz.ch}\\
 \Name{Melanie N. Zeilinger} \Email{mzeilinger@ethz.ch}\\
 \addr Institute for Dynamic Systems and Control, ETH Zurich, Zurich, Switzerland
}

\begin{document}

\maketitle

\begin{abstract}%
 To control a dynamical system it is essential to obtain an accurate estimate of the current system state based on uncertain sensor measurements and existing system knowledge.
 An optimization-based moving horizon estimation (MHE) approach uses a dynamical model of the system, and further allows for integration of physical constraints on system states and uncertainties, to obtain a trajectory of state estimates.
 In this work, we address the problem of state estimation in the case of constrained linear systems with parametric uncertainty.
 The proposed approach makes use of differentiable convex optimization layers to formulate an MHE state estimator for systems with uncertain parameters.
 This formulation allows us to obtain the gradient of a squared and regularized output error, based on sensor measurements and state estimates, with respect to the current belief of the unknown system parameters.
 The parameters within the MHE problem can then be updated online using stochastic gradient descent (SGD) to improve the performance of the MHE.
 In a numerical example of estimating temperatures of a group of manufacturing machines, we show the performance of tuning the unknown system parameters and the benefits of integrating physical state constraints in the MHE formulation.%
\end{abstract}


\section{Introduction}

In order to control complex safety-critical dynamical systems, such as, e.g., flexible and efficient manufacturing systems or power systems, it is necessary to have access to an accurate estimate of the state of the system.
A commonly used state estimation approach is Kalman filtering \citep{Kalman1960}, with an optimal closed-form solution for linear systems with Gaussian disturbances and measurement noise.
An alternative approach is moving horizon estimation (MHE), where a trajectory of state estimates is optimized online to explain the observed outputs with minimal disturbance and noise values, with the last state in the trajectory being the current state estimate~\citep{Rawlings2019}.
In contrast to Kalman filtering, an MHE approach allows us to handle different distributions of the uncertainty and explicitly consider physical constraints on states and uncertainties.
Additionally, MHE is a promising approach, especially for nonlinear state estimation, since it provides robust stability properties, see, e.g.,~\citep{Mueller2017,Schiller2022}.

State estimation algorithms, including MHE, rely on a sufficiently descriptive model of the underlying system.
While it is often possible to derive the model structure of a physical system, finding the true values of system parameters based on noisy measurements is challenging.
Estimating system parameters from input/output data offline is widely studied in the field of system identification, e.g., based on prediction error methods or subspace identification~\citep{Ljung1999}.
However, the use of online sensor measurements to adapt the parameters of the model within an MHE estimator in order to improve the estimation performance is an open research question.
In this work, we propose an MHE approach for simultaneous state estimation and parameter tuning for linear systems subject to bounded disturbances and measurement noise, and physical constraints on the system state.
We rely on the certainty equivalence principle (see, e.g.,~\cite{Tsiamis2019,Dean2020}), where a current belief of the system parameter is used within the MHE problem.
The parameter is thereby initialized through classical system identification, while new measurements are used to improve the belief of the parameter, or adapt it to time-varying changes, in a gradient-based manner.

\textit{Contribution:} An MHE problem is formulated as a differentiable convex optimization layer \citep{Agrawal2019}, allowing for a seamless embedding into efficient machine learning frameworks, such as, e.g., PyTorch~\citep{Paszke2019} or TensorFlow~\citep{Abadi2019}, providing automatic differentiation~(AD) capabilities (see, e.g., \cite{Griewank2008}).
The performance of the MHE estimator using the current parameter belief is evaluated online based on the available input/output data through a squared output error loss function.
The formulation as a convex optimization layer allows us to obtain the gradient of the loss function with respect to the current parameter belief.
This gradient can then be used to update the belief of the system parameter using (projected) stochastic gradient descent (SGD), which under mild assumptions convergences to a local minimizer.
The proposed framework therefore allows for learning-based online improvements of the MHE estimator in a computationally efficient manner.
Relying on existing machine learning frameworks permits a simple implementation in practice and the combination of the MHE layer with additional layers, e.g., a neural network mapping camera images to low dimensional features for pre-processing of sensor measurements, or a convex optimization control policy layer \citep{Agrawal2020} allowing for simultaneous improvements of the shared system parameters within the estimator and controller.
The presented framework allows for an extension to nonlinear MHE using a sequential quadratic programming algorithm similar to \cite{Amos2018}.

\textit{Related work:} Gradient-based parameter updates for discrete-time LTI systems without simultaneous state estimation were addressed within the area of system identification \citep{Bergboer2002}.
Adaptive state observers for combined state and parameter estimation were discussed in~\cite{Ljung1979} and Chapter 10 of~\cite{Anderson1979}.
An expectation-maximization algorithm based on linear Kalman filtering was introduced in \cite{Gibson2005} to learn the parameters of the underlying system model under uncertain measurements.
In the area of MHE, parametric uncertainty was addressed by directly including the parameter estimation in the MHE problem, resulting in a bilinear optimization problem even in the case of linear systems~\citep{Kuhl2011}.
Alternatively, the problem was addressed using a min-max approach, where first the MHE objective was maximized over all possible values of the unknown parameter, and then minimized to find the state estimates \citep{Alessandri2012,Copp2016}.
In the proposed approach, the optimization problem remains convex, and we omit the conservatism introduced by considering the uncertain parameter in a worst-case fashion.
Gradient-based updates of parameters to improve the performance of model predictive control (MPC) were introduced in \cite{Gros2020}, and recently extended to a combined MHE-MPC framework in \cite{Esfahani2021}.
In contrast, the approach presented in this paper permits online improvements of an MHE estimator for constrained linear systems, and the use of existing AD frameworks allows for an efficient and simple implementation.

\textit{Notation:}
A truncated Gaussian distribution of random variable $w\in\mathbb{R}^{n}$ with convex support set $\mathcal{W}\subseteq \mathbb{R}^n$, mean $\bar{w}\in\mathbb{R}^{n}$, and covariance $Q\in\mathbb{R}^{n\times n}$ is denoted as $\mathcal{N}_{\mathcal{W}}(\bar{w},Q)$.
By $\mathbb{E}_w(x)$ we denote the expected value of $x$ with respect to the random variable $w$, by $\left(A\right)_{i,j}$ the element in the $i$-th row and $j$-th column of matrix $A$, by $\left(b\right)_{i}$ the $i$-th element of the column vector $b$, and by~$\mathbb{I}$ the identity matrix of appropriate dimension. The vector $\nicefrac{\partial x}{\partial a} \in \mathbb{R}^n$ contains the partial derivatives of each element of the vector $x \in \mathbb{R}^n$ with respect to the parameter $a\in \mathbb{R}$.
\section{Preliminaries}
\subsection{Problem Formulation}
We consider a linear time-invariant discrete-time system 
\begin{subequations}
	\begin{align}
	x(k+1) &= A(\theta) x(k) + B(\theta) u(k) + w(k), \label{eq:dynamics} \\
	y(k) &= C(\theta) x(k) + v(k), \label{eq:measurement}
	\end{align}	\label{eq:system_model}%
\end{subequations}
with state $x(k)\in \mathbb{R}^n$, input $u(k) \in \mathbb{R}^m$, output $y(k) \in \mathbb{R}^p$, disturbance $w(k)\in\mathbb{R}^{n}$, measurement noise $v(k)\in\mathbb{R}^{p}$, and time index $k \in \mathbb{N}$ with $k\ge0$. The system matrices $A(\theta) \in \mathbb{R}^{n \times n}$, $B(\theta) \in \mathbb{R}^{n \times m}$, and $C(\theta) \in \mathbb{R}^{p \times n}$ depend on a fixed but unknown parameter vector $\theta \in \mathbb{R}^q$ contained within some convex and compact set $\Theta \subset \mathbb{R}^{q}$.
\begin{assumption}\label{ass:distribution}
	The disturbance $w(k)$ and noise $v(k)$ are distributed according to (truncated) Gaussian distributions $\mathcal{N}_{\mathcal{W}}(0,Q)$ and $\mathcal{N}_{\mathcal{V}}(0,R)$, with bounded and convex set $\mathcal{W}\subset \mathbb{R}^n$, and convex set $\mathcal{V}\subseteq \mathbb{R}^p$, both containing the origin in their interior.
\end{assumption}
\begin{assumption}
	The pair $(C(\theta),A(\theta))$ is detectable for any parameter $\theta$ within the set $\Theta$. \label{ass:observability}
\end{assumption}
The system \eqref{eq:system_model} is subject to physical state constraints
\begin{equation}\label{eq:state_constraints}
x(k) \in \mathcal{X} \quad \forall k\ge 0,
\end{equation}
with convex set $\mathcal{X}\subseteq\mathbb{R}^n$.
\begin{assumption}
	The initial state $x(0)\in\mathcal{X}_0 \subseteq \mathcal{X}$ is distributed according to a truncated Gaussian distribution $\mathcal{N}_{\mathcal{X}_0}(\bar{x}_0,P_0)$, with mean $\bar{x}_0\in \mathbb{R}^n$, positive definite covariance matrix $P_0\in\mathbb{R}^{n\times n}$, and bounded and convex set $\mathcal{X}_0 \subset \mathbb{R}^n$.\label{ass:initial_state}%
\end{assumption}
\begin{assumption}\label{ass:constraint_satisfaction}
	System \eqref{eq:system_model} always satisfies the state constraints \eqref{eq:state_constraints}, i.e., at time step $\bar{k}$ an input $u(\bar{k})$ is applied to the system \eqref{eq:dynamics} ensuring $x(k) \in \mathcal{X}$ for all $k > \bar{k}$ and all possible disturbance values $w(k)\in\mathcal{W}$.
\end{assumption}

Assumption \ref{ass:constraint_satisfaction} can, e.g., be satisfied for stable autonomous linear systems \eqref{eq:system_model} with $u(k)=0 \text{ for all } k$, if the initial state $x(0)$ lies within a robust positive invariant set $\Omega$ satisfying state constraints, i.e., $x(0)\in\Omega\subseteq\mathcal{X}$.
Assumption~\ref{ass:constraint_satisfaction} can also be satisfied if a safety control policy $u(k) = \pi_s(k)$ is able to ensure constraint satisfaction. The control policy $\pi_s$ could, e.g., be a human safety pilot, or a safety input $\pi_s$ could be applied whenever a system is coming close to a safety constraint, e.g., detected by a threshold sensor, a low-cost and easy to integrate sensor type (see, e.g., \cite{Alessandri2020}).
Satisfaction of Assumption~\ref{ass:constraint_satisfaction} is not possible for unbounded disturbances.
Therefore, we require the disturbance $w(k)$ to be contained within a bounded set $\mathcal{W}$, while the measurement noise $v(k)$ can potentially be unbounded.

The objective of the presented problem is to obtain an accurate estimate $\hat{x}(k)$ of the state $x(k)$ of the system \eqref{eq:system_model} at each time step $k$, while only having access to noisy sensor measurements $y(k)$, as well as a prior estimate $\bar{x}_0$ of the initial state $x(0)$.
This state estimation problem can be addressed by applying an MHE scheme, which can directly consider the constraints on disturbances and system states during the estimation.
However, an MHE estimator depends on the system model, and thus the unknown parameter $\theta$.
In the following, we propose an iterative approach, where alternately, an MHE based on a belief $\hat{\theta}$ of the unknown parameter is used for state estimation in a certainty equivalent manner (see, e.g., \cite{Tsiamis2019,Dean2020}), and then the state estimates obtained are used to update the parameter belief to improve the performance of the state estimation.
We specify the MHE problem based on the parameter belief in Section~\ref{sec:mhe}, and state the algorithm for iterative parameter updates in Section~\ref{sec:lb_mhe}.

\subsection{Moving Horizon Estimation}\label{sec:mhe}
Given a belief $\hat{\theta}$ of the unknown model parameter $\theta$, the constrained MHE optimization problem with horizon $N$ at time step $k$ can be written as
\begin{subequations}
	\begin{alignat}{2}
	\hat{V}^*_k = \min_{\mathbf{\hat{x}}, \mathbf{\hat{w}}, \mathbf{\hat{v}}} &\left(\| P_{k-N}^{-\nicefrac{1}{2}}(\hat{x}_{k-N|k} - \hat{x}(k-N))\|_2^2 + \sum_{i=k-N}^{k-1}\| Q^{-\nicefrac{1}{2}}\hat{w}_{i|k}\|_2^2 + \|R^{-\nicefrac{1}{2}}\hat{v}_{i|k}\|_2^2\right) \label{eq:constrained_mhe_objective} \\
	\text{s.t. } & \hat{x}_{i+1|k} = A(\hat{\theta})\hat{x}_{i|k} + B(\hat{\theta})u(i) + \hat{w}_{i|k}, &&\hspace*{-4.5cm}\forall i \in \{k-N, \ldots,k-1\}, \\
	& y(i) = C(\hat{\theta})\hat{x}_{i|k} + \hat{v}_{i|k}, &&\hspace*{-4.5cm}\forall i \in \{k-N, \ldots,k-1\}, \\
	& \hat{w}_{i|k} \in \mathcal{W}, \hat{v}_{i|k} \in \mathcal{V}, &&\hspace*{-4.5cm}\forall i \in \{k-N, \ldots,k-1\}, \label{eq:inequality_1} \\
	& \hat{x}_{i|k} \in \mathcal{X}, &&\hspace*{-4.5cm}\forall i \in \{k-N, \ldots,k\}, \label{eq:inequality_2}
	\end{alignat} \label{eq:constrained_mhe_problem}%
\end{subequations}
where $\mathbf{\hat{x}} = \{\hat{x}_{i|k}\}_{i=k-N}^{k}$ is a sequence of state estimates, $\mathbf{\hat{w}} = \{\hat{w}_{i|k} \}_{i=k-N}^{k-1}$ and $\mathbf{\hat{v}} = \{\hat{v}_{i|k} \}_{i=k-N}^{k-1}$ are the corresponding sequences of disturbance and noise estimates, and $\hat{x}(k-N)$ is a past MHE estimate.
The matrices $Q$ and $R$ in the cost are the covariance matrices of the disturbance and measurement noise according to Assumption~\ref{ass:distribution}. The prior weighting $P_{k-N}$ is obtained as
\begin{equation}
	P_{k+1} = Q + A(\hat{\theta})P_kA(\hat{\theta})^{\top} - A(\hat{\theta})P_kC(\hat{\theta})^{\top}(R + C(\hat{\theta})P_kC(\hat{\theta})^{\top})^{-1}C(\hat{\theta})P_kA(\hat{\theta})^{\top}, \label{eq:prior_riccati}
\end{equation}initialized with the covariance matrix of the initial state $P_0$ according to Assumption~\ref{ass:initial_state}.
For time steps $k\le N$, the MHE problem~\eqref{eq:constrained_mhe_problem} is solved with $N=k$.
The problem~\eqref{eq:constrained_mhe_problem} with $N=k$ is called full information estimation problem, in which all available measurements are considered.
MHE is a computationally tractable approximation of full information estimation in which only the past $N$ measurements are considered and the prior weighting in~\eqref{eq:constrained_mhe_objective} with $P_{k-N}$ according to~\eqref{eq:prior_riccati} is used to approximate the effect of the neglected measurements.
The optimal MHE state estimate at time step $k$ is then defined as
\begin{equation}
\hat{x}(k)=\hat{x}^*_{k|k}, \label{eq:state_estimate}
\end{equation}
and the MHE problem~\eqref{eq:constrained_mhe_problem} is solved in a receding horizon fashion.

The motivation for the particular choice of the objective function~\eqref{eq:constrained_mhe_objective} originates from the unconstrained estimation problem with standard Gaussian distributions and no state constraints, i.e., $\mathcal{X,W,X}_0=\mathbb{R}^n,\mathcal{V}=\mathbb{R}^p$, and with known true parameter $\theta$.
In this case, the prior weighting based on~\eqref{eq:prior_riccati} captures the effect of the past measurements which are not included in the MHE horizon, and the solution of the MHE problem is equivalent to the one of the full information estimation problem \citep{Rao2001}, which in this case is the maximum a posteriori sequence of state estimates $\{\hat{x}_{i|k}^*\}_{i=0}^k$ conditioned on all available output measurements \citep{Rao2000}.
While this generally does not hold in the considered constrained case with Assumptions~\ref{ass:distribution},~\ref{ass:initial_state} and~\ref{ass:constraint_satisfaction}, the choice of the objective~\eqref{eq:constrained_mhe_objective} with prior weighting~\eqref{eq:prior_riccati} recovers the full information solution in the case where the inequality constraints~\eqref{eq:inequality_1} and~\eqref{eq:inequality_2} are inactive~\citep{Rao2001}.

We can express the MHE problem \eqref{eq:constrained_mhe_problem} at each time step $k$ as the following mapping from inputs, measurements, and problem parameters to the state estimate $\hat{x}(k)$, i.e.,
\begin{equation}
\hat{x}(k) = \mathrm{MHE}\left(\underbrace{A(\hat{\theta}),B(\hat{\theta}),C(\hat{\theta}),P_{k-N}^{-\nicefrac{1}{2}},\mathbf{y}(k), \mathbf{u}(k), \hat{x}(k-N),Q^{-\nicefrac{1}{2}},R^{-\nicefrac{1}{2}}}_{p}\right), \label{eq:mhe_layer}
\end{equation}
where $\mathbf{y}(k)=\{y(i)\}_{i = k-N}^{k-1}$, $\mathbf{u}(k)=\{u(i)\}_{i=k-N}^{k-1}$, and the list $p$ contains all parameters of~\eqref{eq:constrained_mhe_problem}.

\subsection{Disciplined Parametrized Programming}\label{sec:DPP}
The state estimate \eqref{eq:state_estimate} depends on the parameter belief~$\hat{\theta}$ through the MHE problem~\eqref{eq:constrained_mhe_problem}.
In order to update $\hat{\theta}$ in a gradient-based manner, we want to differentiate the resulting MHE estimate $\hat{x}(k)$ with respect to parameter $\hat{\theta}$.
The disciplined parameterized programming (DPP) grammar \citep{Agrawal2019} allows for the design of convex optimization problems, for which the solution of the problem can be differentiated with respect to its parameters.
A general parameterized program
\begin{subequations}
	\begin{align}
	\min_x & f_0(x,p) \\
	\text{s.t. } & f_i(x,p) \le 0, \text{ } g_i(x,p) = 0,
	\end{align}\label{eq:DPP_program}%
\end{subequations}%
with variables $x\in \mathbb{R}^n$ and list of parameters $p$, is in DPP form, provided the functions $f_i(.,.)$ are convex and $g_i(.,.)$ affine, and both satisfy the DPP grammar.
The following definition provides expressions satisfying the DPP grammar, which are then used in the following to design an MHE problem in DPP form.
\begin{definition}[based on \cite{Agrawal2019}]
	An expression $\phi(x,p)$ satisfies the DPP grammar if it is a linear combination of
	\begin{enumerate}
		\item $Fx$, where the matrix $F$ is a parameter contained in $p$ and the vector $x$ a variable,
		\item $\|F x\|_2^2$, where $F$ is a parameter contained in $p$ or a constant matrix and $x$ a variable.
	\end{enumerate}\label{def:dpp}
\end{definition}

\section{Learning-based Moving Horizon Estimation} \label{sec:lb_mhe}
In this section, we present our proposed approach for MHE state estimation for constrained linear systems with parametric uncertainty.
We start by introducing the online estimator tuning problem used to improve the performance of the MHE estimator by adapting the parameter vector~$\hat{\theta}$ based on current state estimates.
Afterwards, we show how constructing a constrained MHE problem based on the DPP grammar allows us to differentiate the resulting state estimate with respect to the uncertain parameter.
Finally, we show the practical algorithm to update the parameter estimates based on a sampled squared output loss and a stochastic gradient descent method.

\subsection{Online Estimator Tuning for Constrained Systems}

\begin{figure}[t]
	\centering
	\includegraphics[width=0.7\textwidth]{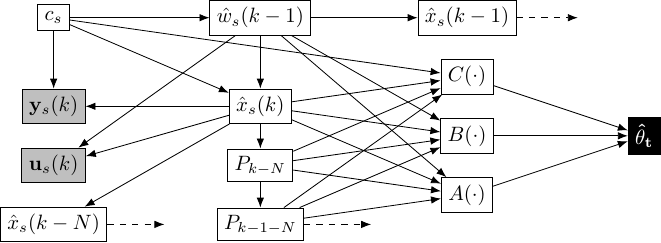}
	\caption{\vspace{-0.2cm}Directed graph showing the construction of each term $c_s$ in~\eqref{eq:c_s}, where each node corresponds to a function (white boxes), fixed parameters (gray boxes) or parameter $\hat{\theta}_t$ (black box), and the edges denote function compositions. The dashed arrows indicated further dependencies on the parameter $\hat{\theta}_t$.} \label{fig:directed_graph}
\end{figure}

In order to improve the performance of the MHE estimator during online estimation we rely on output-error system identification \citep{Verhaegen1992,Bergboer2002}, where a prediction model is used to map a known input sequence to a predicted output sequence.
The parameters of the prediction model are then chosen to minimize a squared norm cost between actual measurements and predicted outputs.
This leads to the following estimator tuning problem
\begin{subequations} %
	\begin{alignat}{2}
	\min_{\hat{\theta}} \text{ } & \mathbb{E}_{w,v,x(0),u}\left[\sum_{k=1}^{n_T}\|y(k) - C(\hat{\theta})\hat{x}(k)\|_2^2 +  \gamma \|\hat{w}(k-1)\|_2^2 \right] \label{eq:tuning_problem_LS_objective} \\
	\text{s.t. } &\hat{x}(k)\!=\!\mathrm{MHE}\left(A(\hat{\theta}),B(\hat{\theta}),C(\hat{\theta}),P_{k-N}^{-\nicefrac{1}{2}},\mathbf{y}(k), \mathbf{u}(k), \hat{x}(k\!-\!N)\right), &&\forall k \in \{1,\ldots,n_T\},\\
	&\hat{w}(k-1) = \hat{x}(k) - A(\hat{\theta})\hat{x}(k-1) - B(\hat{\theta})u(k-1), &&\forall k \in \{1,\ldots,n_T\}, \label{eq:tuning_problem_w_k-1} \\
	&\hat{\theta} \in \Theta, %
	\end{alignat} \label{eq:tuning_problem_LS}%
\end{subequations} %
where the expectation in~\eqref{eq:tuning_problem_LS_objective} is taken over disturbances $w$, measurement noise $v$, initial conditions~$x(0)$, and input sequences $u$.
The second term in the objective \eqref{eq:tuning_problem_LS_objective} with weighting parameter $\gamma \ge 0$ is used to prevent overfitting of the parameter to minimize the loss on the measurements, while the quality of the disturbance estimates deteriorates.
We can not analytically find an optimizer $\hat{\theta}^*$ for the tuning problem~\eqref{eq:tuning_problem_LS}, since the MHE estimator is an implicit mapping from parameter vector $\hat{\theta}$ to state estimate $\hat{x}(k)$ and thus, we cannot directly evaluate the expectation in the objective~\eqref{eq:tuning_problem_LS_objective}.
We therefore turn to an iterative approach, where the objective~\eqref{eq:tuning_problem_LS_objective} is approximated given a parameter belief $\hat{\theta}$ and input/output data starting from $n_S$ initial conditions and running the MHE estimator over $n_T$ time steps as
\begin{align}
\hat{J}(\hat{\theta})&\!=\!\frac{1}{n_S}\sum_{s=1}^{n_S}\sum_{k=1}^{n_T}\left(\|y_s(k)\!-\!C(\hat{\theta})\hat{x}_s(k)\|_2^2\!+\!\gamma\|\hat{w}_s(k\!-\!1)\|_2^2\right) \label{eq:tuning_problem_sampled_objective} \\
&\coloneqq \frac{1}{n_S}\sum_{s=1}^{n_S}\sum_{k=1}^{n_T} c_s\left(y_s(k),C(\hat{\theta}),\hat{x}_s(k),\hat{w}_s(k-1)\right), \label{eq:c_s}
\end{align}
where a sampled input sequence $\{u_s(i)\}_{i=0}^{T-1}$ is applied for each sampled initial condition $x_s(0)$.
The construction of each term $c_s$ in the sampled cost $\hat{J}(\hat{\theta})$~\eqref{eq:tuning_problem_sampled_objective} is depicted in Figure~\ref{fig:directed_graph} as a directed graph.
This sampled loss can be used to update the parameter belief $\hat{\theta}$ in a gradient-based manner using the gradient of the loss $\hat{J}(\hat{\theta})$ with respect to the parameter belief $\hat{\theta}$.
In the following subsection, we show that the MHE problem in DPP form allows us to obtain this cost gradient $\nabla_{\hat{\theta}}\hat{J}(\hat{\theta})$.

\subsection{MHE as Convex Optimization Layer}
The MHE problem~\eqref{eq:constrained_mhe_problem} can be written in DPP form as
\begin{subequations}
	\begin{alignat}{2}
	\min_{\mathbf{\hat{x}}, \mathbf{\hat{w}}, \mathbf{\hat{v}}, \tilde{x}, \tilde{u}} &\left(\|P_{k-N}^{-\nicefrac{1}{2}} \tilde{x}\|_2^{2} + \sum_{i=k-N}^{k-1} \|Q^{-\nicefrac{1}{2}} \hat{w}_{i|k}\|_2^{2} + \|R^{-\nicefrac{1}{2}} \hat{v}_{i|k}\|_2^{2}\right) \label{eq:DPP_mhe_problem_objective}\\
	\text{s.t. } & \hat{x}_{i+1|k} = A(\hat{\theta})\hat{x}_{i|k} + B(\hat{\theta})\tilde{u}_i + \hat{w}_{i|k}, &&\hspace*{-2.5cm}\forall i \in \{k-N, \ldots,k-1\}, \\
	& y(i) = C(\hat{\theta})\hat{x}_{i|k} + \hat{v}_{i|k}, &&\hspace*{-2.5cm}\forall i \in \{k-N, \ldots,k-1\},\\
	& \hat{w}_{i|k} \in \mathcal{W}, \hat{v}_{i|k} \in \mathcal{V}, &&\hspace*{-2.5cm}\forall i \in \{k-N, \ldots,k-1\},\\
	& \hat{x}_{i|k} \in \mathcal{X}, &&\hspace*{-2.5cm}\forall i \in \{k-N, \ldots,k\},\\
	& \tilde{u}_i = u(i), &&\hspace*{-2.5cm}\forall i \in \{k-N, \ldots,k-1\}, \label{eq:DPP_mhe_problem_aux_inputs} \\
	& \tilde{x} = \hat{x}_{k-N|k} - \hat{x}(k-N), \label{eq:DPP_mhe_problem_aux_initial}
	\end{alignat} \label{eq:DPP_mhe_problem}%
\end{subequations}
where all terms satisfy the conditions in Definition~\ref{def:dpp}.
Since both $P_{k-N}^{-\nicefrac{1}{2}}$ and $\hat{x}(k-N)$ are parameters of~\eqref{eq:mhe_layer}, the auxiliary variable $\tilde{x}$ is introduced in~\eqref{eq:DPP_mhe_problem_aux_initial}, such that the prior weighting $\|P_{k-N}^{-\nicefrac{1}{2}} \tilde{x}\|_2^{2}$ satisfies condition $2)$ of Definition~\ref{def:dpp}.
Similarly, introducing the auxiliary variables $\tilde{u}_i$ in~\eqref{eq:DPP_mhe_problem_aux_inputs} ensures that the terms $B(\hat{\theta})\tilde{u}_i$ satisfy condition $1)$ of Definition~\ref{def:dpp}.
The formulation~\eqref{eq:DPP_mhe_problem} allows us to differentiate the MHE state estimate \eqref{eq:state_estimate} with respect to each element of the list of parameters $p$.
Specifically, since the parameters $A(\hat{\theta})$, $B(\hat{\theta})$, $C(\hat{\theta})$, $P_{k-N}^{-\nicefrac{1}{2}}$ and $\hat{x}(k-N)$ depend on the parameter belief $\hat{\theta}$, the partial derivative of $\hat{x}(k)$ with respect to each element of $\hat{\theta}$ can be obtained based on the chain rule as
\begin{equation}
\begin{split}
	\frac{\partial \hat{x}(k)}{\partial \left(\hat{\theta}\right)_j} =& \sum_{s=1}^{n}\sum_{l = 1}^{n}\frac{\partial \mathrm{MHE}(\cdot)}{\partial \left(A(\hat{\theta})\right)_{s,l}} \frac{\partial \left(A(\hat{\theta})\right)_{s,l}}{\partial \left(\hat{\theta}\right)_j} + \sum_{s=1}^{n}\sum_{l = 1}^{m}\frac{\partial \mathrm{MHE}(\cdot)}{\partial \left(B(\hat{\theta})\right)_{s,l}} \frac{\partial \left(B(\hat{\theta})\right)_{s,l}}{\partial \left(\hat{\theta}\right)_j} \\
	&+ \sum_{s=1}^{p}\sum_{l = 1}^{n}\frac{\partial \mathrm{MHE}(\cdot)}{\partial \left(C(\hat{\theta})\right)_{s,l}} \frac{\partial \left(C(\hat{\theta})\right)_{s,l}}{\partial \left(\hat{\theta}\right)_j} + \sum_{s= 1}^{n}\sum_{l = 1}^{n}\frac{\partial \mathrm{MHE}(\cdot)}{\partial \left(P_{k-N}^{-\nicefrac{1}{2}}\right)_{s,l}} \frac{\partial \left(P_{k-N}^{-\nicefrac{1}{2}}\right)_{s,l}}{\partial \left(\hat{\theta}\right)_j} \\
	&+ \sum_{s=1}^{n}\frac{\partial \mathrm{MHE}(\cdot)}{\partial \left(\hat{x}(k-N)\right)_s}\frac{\partial \left(\hat{x}(k-N)\right)_s}{\partial \left(\hat{\theta}\right)_j}\label{eq:estimate_partial}. \\
\end{split}
\end{equation}
Thereby, the partial derivatives of the MHE estimator~\eqref{eq:mhe_layer} with respect to all elements of $A(\hat{\theta})$, $B(\hat{\theta})$, $C(\hat{\theta})$, $P_{k-N}^{-\nicefrac{1}{2}}$, and $\hat{x}(k-N)$ can be obtained based on the implementation of differentiable optimization layers \citep{Agrawal2019}, and the partial derivatives $\nicefrac{\partial \left(P_{k-N}^{-\nicefrac{1}{2}}\right)_{s,l}}{\partial \left(\hat{\theta}\right)_j}$ and $\nicefrac{\partial \left(\hat{x}(k-N)\right)_s}{\partial \left(\hat{\theta}\right)_j}$ can be obtained recursively through AD based on, e.g., PyTorch \citep{Paszke2019}.
\begin{remark}
	Due to the selected prior weighting, choosing the MHE horizon length $N=1$ recovers the standard linear Kalman filter (KF) in the unconstrained case \citep{Rao2001}.
	Therefore, the proposed formulation also allows for AD of the KF with respect to a belief of the system parameters and for updating these parameters based on stochastic gradient descent.
	Alternatively, a KF can also be implemented directly in PyTorch~\citep{Paszke2019} or TensorFlow~\citep{Abadi2019} to obtain the derivative of state estimates with respect to a current parameter belief. \label{re:KF_update}
\end{remark}

\begin{remark}
		The computational complexity of the proposed framework can be reduced by assuming observability (instead of detectability) in Assumption~\ref{ass:observability}.
		Consequently, an MHE can be formulated without prior weighting, resulting in the last two terms in~\eqref{eq:estimate_partial} to vanish, and thus the differentiation through the Riccati iteration~\eqref{eq:prior_riccati} is no longer necessary.
\end{remark}

\subsection{Gradient-based Update of Model Parameters}
Given an initial belief $\hat{\theta}_0$ of the unknown parameter, we use a projected stochastic gradient method \citep{Kushner1978} to update our belief of the unknown parameter in the MHE estimator.
Thereby, in each learning epoch $t \in \mathbb{N}$, with $t \ge 0$, we sample the loss \eqref{eq:tuning_problem_sampled_objective} and obtain the gradient of the sampled loss with respect to the current parameter value, i.e., $\nabla_{\hat{\theta}_t}\hat{J}(\hat{\theta}_t)$.
This gradient can be obtained by applying the chain rule along the graph shown in Figure~\ref{fig:directed_graph} for each term $c_s$ in~\eqref{eq:c_s}, which requires the partial derivatives of the state estimates $\hat{x}(k)$ and $\hat{x}(k-1)$ with respect to the parameter belief $\hat{\theta}$ as obtained in~\eqref{eq:estimate_partial}.
The overall gradient of the loss can be computed through AD.
The parameter belief $\hat{\theta}$ is then updated as
\begin{equation}
\hat{\theta}_{t+1} = \Pi_{\Theta}(\hat{\theta}_t - \alpha_t \nabla_{\hat{\theta}_t}\hat{J}(\hat{\theta}_t)) \label{eq:projected_SGD}
\end{equation}
where $\Pi_{\Theta}$ is the projection onto the set $\Theta$ and $\alpha_t$ is a learning rate satisfying
\begin{equation}
\sum_{t=0}^{\infty}\alpha_t^2 < \infty , \quad \sum_{t=0}^{\infty}\alpha_t = \infty. \label{eq:learning_rate_condition}
\end{equation}
A learning rate which leads to good practical convergence for unconstrained SGD is, e.g., given in \cite{Goldstein1988} as  
\begin{equation}
\alpha_t = \frac{\alpha_0}{t} \quad \forall t \ge 1,\label{eq:lr_update}
\end{equation}
with $\alpha_0>0$.
The online adaption of the parameter is summarized in Algorithm~\ref{alg:learning_mhe_parameters}, where we alternately sample the loss \eqref{eq:tuning_problem_sampled_objective} for $n_S$ initial conditions and run the MHE estimator over $n_T$ time steps each, and then update the parameter belief based on \eqref{eq:projected_SGD}.

Convergence of projected SGD is shown for problems with differentiable convex objective functions and compact convex projection sets, provided some assumptions on the learning rate are satisfied \citep{Kushner1978,Cohen2017}.
In the non-convex case presented here, we expect at least convergence to a local minimizer of the estimator tuning problem~\eqref{eq:tuning_problem_LS}.

\begin{algorithm}[t]
	\caption{Online learning of MHE parameters.}
	\begin{algorithmic}[1]
		\REQUIRE Initial parameter belief $\hat{\theta}_0$, MHE layer $\mathrm{MHE}(\cdot)$, initial learning rate $\alpha_0$
		\ENSURE $\hat{\theta}$
		\WHILE{$\hat{\theta}_{t+1}$ not equal to $\hat{\theta}_t$}
		\STATE initialize loss to zero, i.e., $\hat{J}(\hat{\theta}_t) = 0$, and sample $n_S$ initial conditions $x_0 \sim \mathcal{N}_{\mathcal{X}_0}(\bar{x}_0,P_0)$ 
		\FOR{every sample $s = 1,2,\ldots, n_S$}
		\STATE initialize sample loss to zero, i.e., $\hat{J}_S(\hat{\theta}_t) = 0$
		\FOR{every time step $k = 1,2,\ldots, n_T$}
		\STATE sample an input $u(k-1)$ ensuring $x(k) \in \mathcal{X}$ for all disturbances $w(k-1) \in \mathcal{W}$
		\STATE run system \eqref{eq:dynamics}, and obtain sensor measurement $y(k)$ from \eqref{eq:measurement}
		\STATE obtain state estimate $\hat{x}(k)$ solving \eqref{eq:DPP_mhe_problem}
		\STATE update the sample loss $\hat{J}_S(\hat{\theta}_t)$ with the squared output error and regularization
		\ENDFOR
		\STATE update the approximated loss $\hat{J}(\hat{\theta}_t) += \hat{J}_S(\hat{\theta}_t)$
		\ENDFOR
		\STATE obtain gradient of loss $\nabla_{\hat{\theta}_t} \hat{J}(\hat{\theta}_t)$ through AD, and update the parameter belief $\hat{\theta}_{t+1}$ according to \eqref{eq:projected_SGD}
		\STATE update learning rate $\alpha_{t+1}$ according to \eqref{eq:lr_update}
		\ENDWHILE
	\end{algorithmic} \label{alg:learning_mhe_parameters}
\end{algorithm}
\section{Numerical Example}
To demonstrate the efficiency of the presented MHE framework for combined online state estimation and parameter tuning, we consider a cooling system for multiple manufacturing machines in a factory hall.
The manufacturing machines are heating up due to randomly varying production loads, and the temperature of each machine is influencing the temperature of neighboring machines.
We assume the coupling dynamics of the temperatures between subsystems to be a priori unknown and that there exists some form of safety controller able to prevent the temperature of each subsystem to violate a safety critical upper temperature constraint.
Our proposed MHE approach with horizon length $N=10$ is applied to continuously estimate the temperature of each machine, while the belief of the unknown temperature coupling parameter is updated based on newly available measurements.
We use PyTorch \citep{Paszke2019}, the CvxpyLayers package \citep{Agrawal2019}, and pytorch-sqrtm \citep{Li} to differentiate through matrix square-roots.
The code of the numerical example is available at \href{https://github.com/IntelligentControlSystems/mhe-layers/}{https://github.com/IntelligentControlSystems/mhe-layers/}.

We consider a system consisting of $4$ machines arranged in a square order.
The dynamics of the temperatures $T_i(k)$ with $i \in \left\{1,2,3,4\right\}$ of all machines is described by
\begin{equation}
x(k+1) = \left(\mathbb{I} + \frac{T_s}{1000}\begin{bmatrix}
5 & \theta & \theta & 0 \\
\theta & 5 & 0 & \theta \\
\theta & 0 & 5 & \theta \\
0 & \theta & \theta & 5
\end{bmatrix} \right) x(k) - T_su(k) + w(k),
\end{equation}
where $x(k) = \left[T_1(k),T_2(k),T_3(k),T_4(k)\right]^\top$, $u(k) \in \mathbb{R}^4$, $w(k)\sim\mathcal{N}_{\left\{w|\|w\|_\infty \le 0.1\right\}}(0,0.01\mathbb{I})$, the true underlying system parameter is $\theta = 1$, and the sampling time is $T_s = 0.1$.
The local cooling inputs to each machine at all time steps $k$ are constrained to $u_i(k) \in \left\{u|0\le u \le u_{\max}\right\}$ with $u_{\max}=4$. 
The initial state $x(0)$ is distributed according to a truncated Gaussian distribution with mean $\bar{x}_0 = \left[100^\circ C,100^\circ C,100^\circ C,100^\circ C\right]^{\top}$, variance $P_0=\mathbb{I}$, and support set $\mathcal{X}_0 = \{x|(x)_i\le 103^\circ C, \forall i\in\{1,2,3,4\}\}$.
Two temperature sensors are placed such that they measure
\begin{equation}
y(k) = \frac{1}{3} \begin{bmatrix}
1 & 1 & 1 & 0 \\
0 & 1 & 1 & 1
\end{bmatrix} x(k) + v(k)
\end{equation}
with $y(k) \in \mathbb{R}^2$, and $v(k)\sim\mathcal{N}(0,0.1\mathbb{I})$.
The cooling inputs are chosen as sampled sinusoidal trajectories, adjusted by safety control inputs based on threshold sensor measurements.
We run Algorithm~\ref{alg:learning_mhe_parameters} initialized with $\hat{\theta}_0=10$ and with $\Theta=\left\{\theta| 0.1 \le \theta \le 50\right\}$.
Further details on the simulation example can be found in Appendix~\ref{app:numerical_example}.
\begin{figure}[t]
	\centering
	\includegraphics[width=\textwidth]{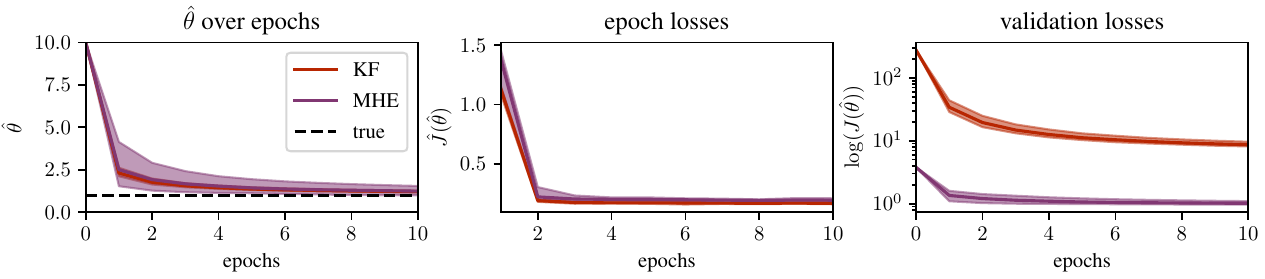}
	\caption{\vspace{-0.9cm}Learning of the parameter $\theta$ over 10 epochs for both our MHE approach and a standard linear Kalman filter.
		The first subplot shows the evolution of the parameter belief $\hat{\theta}$, the second subplot the sampled losses $\hat{J}(\hat{\theta})$ as defined in \eqref{eq:tuning_problem_sampled_objective} and the third subplot the validation loss~$J(\hat{\theta})$.
		The shaded area shows the minimal and maximal values over $20$ different learning instances, while the solid line is the median over all instances.} \label{fig:plot_learning}
\end{figure}
\begin{figure}[t]
	\centering
	\includegraphics[width=\textwidth]{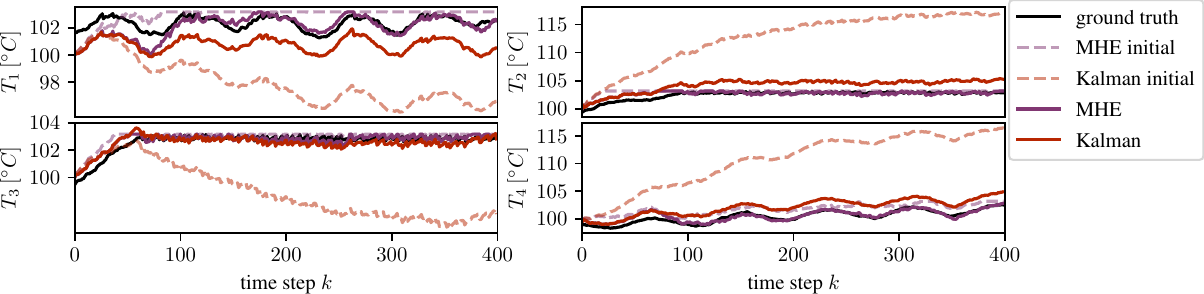}
	\caption{\vspace{-0.9cm}Temperature estimation for both the MHE and Kalman filter approach, with the initial parameter $\hat{\theta}_0$ (dashed lines), and with the learned parameter after 10 learning epochs.} \label{fig:plot_validation}
\end{figure}

In Figure~\ref{fig:plot_learning}, we plot the change of the parameter belief $\hat{\theta}$ over the epochs, as well as the epoch losses according to \eqref{eq:tuning_problem_sampled_objective} and the validation losses, which are the averaged norm distance between the true system state and the state estimates (in logarithmic scale) for both our MHE approach and a standard linear Kalman filter.
The parameters in the Kalman filter are updated in a gradient-based fashion, as outlined in Remark~\ref{re:KF_update}.
In Figure~\ref{fig:plot_validation}, we plot a validation example of the temperature estimation with both our MHE approach and the linear Kalman filter.
While the Kalman filter estimate diverges from the true temperature values initially, after recovering the true underlying parameter, the estimates are closer to the true values.
In comparison, the integration of an upper temperature constraint in the MHE formulation already allows us to provide improved estimates based on the wrong initial parameter and even more after the parameter belief converges to the true parameter.
This also explains the large difference between the validation losses of the MHE approach and the Kalman filter in the third subplot of Figure~\ref{fig:plot_learning}.

Note that we are comparing our MHE approach to a standard unconstrained Kalman filter.
The Kalman filter based estimates could be improved by using a clipping or projection mechanism to ensure that the resulting state estimate satisfies the constraints (see, e.g., \cite{Amor2018} for a review).
It is, however, not straightforward to obtain the gradient of the state estimate with respect to the parameters after applying such a mechanism.

\acks{This work was supported by the Swiss National Science Foundation under grant no. PP00P2\textunderscore157601/1. Simon Muntwiler's research was supported by funds from the Bosch Research Foundation im Stifterverband.}

\bibliography{bibliography}

\appendix

\section{Further Details on the Numerical Example}\label{app:numerical_example}
We assume there to be a threshold temperature sensor available for each machine.
The output of these threshold sensors is
\begin{equation}
y_i^{\text{th}}(k) = \begin{cases}
1 & T_i(k) > T^{\text{th}}, \\
0 & \text{otherwise},
\end{cases}
\end{equation}
with threshold temperature $T^{\text{th}} = 103^\circ \text{C}$.
Based on the threshold information, a safety control law for the cooling input to each machine $i$ is designed as
\begin{equation}
u_i(k) = \begin{cases}
u_{\text{safety}} & y_i^{\text{th}} = 1 \\
u_{i,\text{c}}(k) & \text{otherwise.}
\end{cases}
\end{equation}
where $u_{\text{safety}}$ is a safety cooling input chosen as the maximal input $u_{max}$ here and $u_{i,\text{c}}(k)$ is a proposed cooling input to machine $i$ at time step $k$.
The inputs $u_{i,\text{c}}(k)$ are sampled sinusoidal trajectories.

The loss~\eqref{eq:tuning_problem_sampled_objective} is sampled in each learning epoch by estimating the state of the system starting from $n_S=5$ different initial conditions and simulated over $n_T=400$ time steps with weighting parameter $\gamma = 0.1$.
The parameters $n_S$ and $n_T$ where chosen to trade-off performance of the estimator tuning problem and complexity of the underlying computations.
The learning rate in~\eqref{eq:lr_update} is initialized with $\alpha_0=6$.

\end{document}